\documentclass[a4paper,3p]{article}
\usepackage{lineno}
\modulolinenumbers[5]
\usepackage{hyperref}
\usepackage{graphicx,caption}
\usepackage{cmap}
\usepackage[T2A]{fontenc}
\usepackage[utf8]{inputenc}
\usepackage[russian, english]{babel}
\usepackage{textcomp}
\usepackage{xcolor}
\usepackage{amsmath}
\usepackage{amssymb}
\usepackage{bm}
\usepackage[normalem]{ulem}
\usepackage{physics}
\usepackage{multirow}
\usepackage{hhline}
\usepackage{subfigure}
\usepackage{comment}
\usepackage{float}
\usepackage{multicol}
\usepackage{blindtext}

\begin{document}

	\begin{center}
			{\Large Modeling of shock wave passage through porous copper using moving window technique and kernel gradient correction in smoothed particle hydrodynamics method
			} 
	\end{center}
	\begin{center}
			G.~D.~Rublev$^{1,2}$, S.~A.~Murzov$^{1,2}$
	\end{center}
	\begin{center}
		{\small $^{1)}$Dukhov  Research Institute of Automatics, Russia, Moscow, 127030, Sushchevskaya St., 22}
	\end{center}
	\begin{center}
		{\small $^{2)}$ Joint Institute for High Temperatures of RAS, Moscow, 125412, Russia}
	\end{center}

This paper introduces a novel methodology for modeling stationary shock waves in porous materials, which employs the recently developed moving window technique. The core of this method is the iterative adjustment of the reference frame to the boundary conditions that regulate the entry and exit of Lagrangian particles from a fixed computational domain, which are used to model the flow of a compressible medium. A Godunov-type smoothed particle hydrodynamics (SPH) method with reconstruction of values at the contact is employed for the purposes of modeling. Kernel gradient correction for this method  is proposed to enhance the precision of the approximation. The shock Hugoniot of porous copper is calculated, and the structure of the compacting wave and elastic precursor in porous copper at shock amplitude near the yield strength of solid copper is studied.


\section*{Introduction}
In numerical modeling of hydrodynamic processes associated with the movement of an object over a considerable distance (e.g., in the problem of the motion of a body in a gas or the propagation of a compression impulse in a heterogeneous medium), it is often the case that only processes occurring in a localized region of space surrounding the phenomenon or object under study are of interest. In such a case, modeling the entire computational domain at each moment in time is a computationally costly and impractical approach from a physical standpoint, as the influence of distant computational domain regions on the phenomenon under study is typically insignificant. The influence of distant areas can be effectively mitigated through the precise calibration of boundary conditions on the sides of a smaller rectangular area encompassing the phenomenon under study. The application of the inverse motion method to a rectangular region, in conjunction with Galilean transformations and the combination of the region with a moving coordinate system, allows us to consider the selected rectangle as a ``moving observation window'', with the setting of boundary conditions on each side of the rectangle. 

The most straightforward observation window is a well-established concept in fluid dynamics. Such an observation window is employed, for instance, in the modeling of the flow around an obstacle by a stationary flow of gas or liquid. The realization of such an observation window is contingent upon the correct specification of the boundary conditions at the boundaries of the region and on the surface of the obstacle~\cite{Kryukov:2022, Zhakhovskii:1997, Zhakhovsky:1999}.     

There are, however, a number of urgent problems that cannot be resolved without the use of more sophisticated and complex mowing window techniques. For example, when a shock wave (SW) propagates in an elastic-plastic heterogeneous medium with collapsing pores, the stress relaxation zone behind the SW front will be unsteady in the case of a chaotic arrangement of pores of different sizes, which again requires a study on samples of large length and with large computational costs.  
The problem of SW propagation in porous metals is actively investigated~\cite{Cotton2018, ma15238501,Boade, Marsh1980, Murzov2024}, including by the molecular dynamics method~\cite{doi:10.1063/1.4818487} in the case of vacuumized pores of nanometric size. For larger pore sizes, the molecular dynamics method is not applicable due to the excessively high computational cost associated with the need to have a large number of atoms in the observation window. Our work presents simulation results for relatively weak SWs, where the compactification wave moves much slower than the elastic compression wave, similar to those investigated in~\cite{Boade, Cotton2018}.

The use of Godunov-type smoothed particle hydrodynamics methods~\cite{Parshikov:JCP:2002,MedinParshikov:2010,XiaoyanHu2016,
Zhang2022} allows us to apply a mesomechanical model with explicit material structure to study multi-material media, including porous metals at arbitrary pore size. We use 2D modeling where the Wendland~C$^2$ \cite{Wendland:1995} smoothing function is used as the smoothing kernel.
In 1999, the solution of the Riemann problem on contacts between particles (contact SPH, CSPH)~\cite{Parshikov1999} was introduced into the SPH equations for a compressible nonviscous nonthermal conductive medium, similar to the calculation of fluxes across cell boundaries in the Godunov mesh-based methods~\cite{Toro2013}. The equations of the Godunov-type SPH contact method include momentum and energy fluxes in the right-hand sides of the equations. These fluxes are computed using an approximate solution of the Riemann problem at the contact between the base particle and each of the particles in its neighborhood. The CSPH method uses piecewise constant functions to transfer the values of physical quantities to the point of contact. This method leads to high numerical diffusion, which is the reason of smoothing of SW fronts and fast damping of oscillations in the medium~\cite{Parshikov2023}.
Several approaches have been proposed by different authors to reduce the numerical diffusion, among which it is worth mentioning the use of increased reconstruction order (MUSCL (Monotonic Upstream-centered Scheme for Conservation Laws), WENO (Weighted Essentially Non-Oscillatory))~\cite{LI2022111106,Zhang2022,XiaoyanHu2016}. 

Besides diffusion effects in SPH-schemes there are also errors connected with approximation of differential operators standing in the right parts of initial equations.  To eliminate such effects in the work~\cite{Randles1996} proposed to use the Taylor series expansion of the functions standing in the SPH sums by introducing a corrective renormalization matrix into the equations. This approach has found wide application in works by other authors (see, e.g.,~\cite{Oger2007, Lind2012, Sun2021}). In section~\ref{sec:TKC-MUSCL-SPH}, a new renormalization scheme for elastic-plastic media in Godunov-type SPH methods is presented.

All calculations in the paper were performed using a parallel software package~\cite{EgorovaDPZ19,Dyachkov2017,Murzov2024} developed with the authors' participation. Visualization of 2D distributions of material flow characteristics was obtained using the ParaView~\cite{paraview} program.

\section{The mathematical model}
A system of equations expressing the laws of conservation of mass, momentum, and energy is used: 
\begin{equation}\label{eq:continuity}
	\frac{d\varepsilon}{dt} = \nabla \cdot \overrightarrow{U},
\end{equation}
\begin{equation}\label{eq:momentum}
	\frac{d\overrightarrow{U}}{dt}=\frac{1}{\rho}\nabla \cdot \hat{\sigma},
\end{equation}
\begin{equation}\label{eq:energy}
	\frac{dE}{dt}=-\frac{1}{\rho}\nabla \cdot \left(\hat{\sigma} \cdot \overrightarrow{U}\right),
\end{equation}
where $\varepsilon = -\ln \left(\rho/\rho_0\right)$ is volumetric strain,  $\hat{\sigma}$ is stress tensor, which is represented as the sum of the spherical $-P\hat{I}$ ($\hat{I}$ is a unit tensor) and deviatoric $\hat{S}$ parts:
$$ \hat{\sigma} = -P\hat{I} + \hat{S}. $$
The stress deviator in an elastic medium is calculated as:
\begin{equation}\label{eq:dSdt}
	\frac{d\hat{S}}{dt}= 2G\left(\frac{d\hat{\varepsilon}}{dt} - \frac{1}{3}\text{tr}\left(\frac{d\hat{\varepsilon}}{dt}\right)\hat{I}\right),
\end{equation}
where $G$ is the shear modulus and $\frac{d\hat{\varepsilon}}{dt}$ is the strain rate tensor:
\begin{equation}\label{eq:strainRateDeviator}
	\frac{d\hat{\varepsilon}}{dt} = \frac{1}{2}\left(\nabla\otimes\overrightarrow{U} + \left(\nabla\otimes\overrightarrow{U}\right)^\text{T} \right).
\end{equation}

The von Mises model with constant yield strength is used to describe plastic deformations.

The system (\ref{eq:continuity})-(\ref{eq:strainRateDeviator}) is closed by the equation of state in Mie--Grüneisen form
$$P = P_{ref} + \gamma\rho(e - e_{ref})$$
 with two reference curves:
 \begin{equation}
 	P_{ref} = \begin{cases}
 		\rho_0 c_a^2\frac{1-x}{1-s_a(1-x)}, \, x\leq 1,
 		\\
 		\rho_0 c_a^2(1-x), \, x >1,
 	\end{cases}
 \end{equation}
\begin{equation}
	e_{ref} = \begin{cases}
		\frac{1}{2}\left[\frac{c_a(1-x)}{1-s_a(1-x)}\right]^2, \, x\leq 1,
		\\
		c_a^2(1-x)^2/2, \, x >1,
	\end{cases}
\end{equation}
where $x = \rho_0/\rho$. The relation between the SW velocity $u_s$ and the substance velocity jump at the shock $u_p$ is assumed to be linear: $u_s = c_a + s_au_p$. To calculate the interparticle interaction in the Godunov-type SPH method, the longitudinal and transverse speeds of sound must be given. The longitudinal sound speed $c^l$ is determined from the quasi-acoustic approximation~\cite{Bushman:1988}:
\begin{equation}\label{eq:cl}
c^l = \sqrt{c_b^2+4G/(3\rho)},
\end{equation}
where $c_b = \sqrt{B/\rho}$, $B=\rho_0 c_a^2$ is the bulk compression modulus. The transverse speed of sound is $c^t = \sqrt{G/\rho}$ at constant shear modulus $G$.

\section{Modeling technique}\label{sec:TKC-MUSCL-SPH}
In order to improve the approximation accuracy and reduce the numerical diffusion compared to the standard contact SPH (CSPH) method \cite{Parshikov:JCP:2002}, a Godunov-type SPH method with MUSCL-type reconstruction at the contact between particles and kernel correction (Total Kernel Correction Monotonic Upstream-centered Scheme for Conservation Laws Smoothed Particle Hydrodynamics (TKC-MUSCL-SPH) method) was developed and applied. The equations of the modified method are as follows:
\begin{equation}
	\label{SPH-continuity}
	\frac{d\varepsilon_i}{dt}=2\sum\limits_j \frac{m_j}{\rho_j}\left(\overrightarrow{U_{ij}^{*}} - \overrightarrow{U_i}\right)\cdot\left(\mathbf{L}_i^{-1}\cdot\nabla_iW_{ij}\right),
\end{equation}
\begin{equation}
	\label{SPH-momentum}
	\frac{d\overrightarrow{U}_i}{dt} = -2\sum\limits_j \frac{m_j \overrightarrow{\tilde{\sigma}^{*R}_{ij}} }{\rho_j\rho_ih_{ij}}W'_{ij},
\end{equation}
\begin{equation}
	\label{SPH-energy}
	\frac{dE_i}{dt}=-2\sum\limits_j \frac{m_j\overrightarrow{\tilde{\sigma}^{*R}_{ij}}\cdot\overrightarrow{U_{ij}^{*}}}{\rho_j\rho_ih_{ij}}W'_{ij},
\end{equation}
where $W_{ij} = W(q_{ij}|h_{ij})$ is the smoothing kernel, which is a function of $q_{ij} = \left|\left|\overrightarrow{r}_i - \overrightarrow{r}_j\right|\right|/h_{ij}$. The smoothing kernel function depends on the smoothing length $h_{ij} = \theta (D_i + D_j)$ as a parameter. Here $\theta$ is the smoothing length parameter, which is usually chosen to be $0.5-0.7$. The matrix $\mathbf{L}_i$~\cite{Randles1996} is defined by the following equation
\begin{equation}
	\label{L-matrix}
	\mathbf{L}_i = \sum\limits_j \frac{m_j}{\rho_j}\nabla_i W_{ij}\otimes(\vec{r_j} - \vec{r_i}).
\end{equation}
 
When evaluating solution of the the Riemann problem for a pair of particles, the values of the functions $\{\sigma^{\alpha\beta},\rho,c^l, c^t,U^{\alpha}\}=\Phi$ on the contact plane for the ``left'' $\Phi_l$ and ``right'' $\Phi_r$ particles are evaluated using interpolation  
\begin{equation}\label{muscl-reconstruction-scheme}
	\Phi_l = \Phi_i + \frac{1}{2}\delta\Phi_i, \quad
	\Phi_r = \Phi_j - \frac{1}{2}\delta\Phi_j.
\end{equation}
The values of $\delta \Phi_i$ and $\delta \Phi_j$ are calculated using the slope limiter ``$\mathrm{minmod}$'':
\begin{equation}\label{delta_a}
	\delta \Phi_i = \mathrm{minmod}(\Phi_j - \Phi_i, \nabla\Phi_i\cdot\overrightarrow{r_{ji}}),
\end{equation}
\begin{equation}\label{delta_b}
	\delta \Phi_j = \mathrm{minmod}(\nabla\Phi_j\cdot\overrightarrow{r_{ji}}, \Phi_j - \Phi_i).
\end{equation}

The calculation of the values at the contact $\overrightarrow{U}^{*}_{ij}$ and $\overrightarrow{\tilde{\sigma}^{*R}}_{ij}$ is based on their estimation in the acoustic approximation:
\begin{equation*}
	\overrightarrow{U}^{*}_{ij} = M_{RST\rightarrow XYZ}\cdot
	\begin{pmatrix}
		\frac{U_l^{R}\rho_lc^l_l + U_r^{R}\rho_rc^l_r - \tilde{\sigma}^{RR}_r + \tilde{\sigma}^{RR}_l}{\rho_lc^l_l + \rho_rc^l_r}
		\\
	\frac{U_l^{S}\rho_lc^t_l + U_r^{S}\rho_rc^t_r - \tilde{\sigma}^{SR}_r + \tilde{\sigma}^{SR}_l}{\rho_lc^t_l + \rho_rc^t_r} 		
		\\
		\frac{U_l^{T}\rho_lc^t_l + U_r^{T}\rho_rc^t_r - \tilde{\sigma}^{TR}_r + \tilde{\sigma}^{TR}_l}{\rho_lc^t_l + \rho_rc^t_r} 
	\end{pmatrix},	
\end{equation*}

\begin{equation*}
	\overrightarrow{\tilde{\sigma}^{*R}}_{ij} = 
	M_{RST\rightarrow XYZ}\cdot\begin{pmatrix}
		\frac{\tilde{\sigma}^{RR}_r\rho_lc_l^l + \tilde{\sigma}^{RR}_l\rho_rc_r^l - \rho_lc_l^l\rho_rc_r^l\left( U_{r}^R - U_{l}^R \right)}{\rho_lc_l^l  + \rho_rc_r^l}
		\\
		\frac{\tilde{\sigma}^{SR}_r\rho_lc_l^t + \tilde{\sigma}^{SR}_l\rho_rc_r^t - \rho_lc_l^t\rho_rc_r^t\left( U_{r}^S - U_{l}^S \right)}{\rho_lc_l^t  + \rho_rc_r^t}
		\\
		\frac{\tilde{\sigma}^{TR}_r\rho_lc_l^t + \tilde{\sigma}^{TR}_l\rho_rc_r^t - \rho_lc_l^t\rho_rc_r^t\left( U_{r}^T - U_{l}^T \right)}{\rho_lc_l^t  + \rho_rc_r^t}
	\end{pmatrix},	
\end{equation*}
where the stress vector is corrected by the symmetric combination of renormalization matrices:
$$\begin{pmatrix}
\tilde{\sigma}^{RR}_i
\\
\tilde{\sigma}^{SR}_i
\\
\tilde{\sigma}^{TR}_i
\end{pmatrix} = M_{XYZ\rightarrow RST} \cdot\left(\hat{\sigma}_i\cdot \left(\frac{\mathbf{L}_i^{-1} + \mathbf{L}_j^{-1}}{2}\cdot\overrightarrow{e}^R_{ij}\right)\right),$$
$$\begin{pmatrix}
	U^{R}_i
	\\
	U^{S}_i
	\\
	U^{T}_i
\end{pmatrix} = M_{XYZ\rightarrow RST} \cdot \begin{pmatrix}
U^{x}_i
\\
U^{y}_i
\\
U^{z}_i
\end{pmatrix}.
$$
In this context, the notation $M_{XYZ\rightarrow RST}$ represents the transition matrix from the coordinate system $XYZ$ to the local coordinate system $RST$. The unit orthants of this coordinate system are oriented as follows: $\overrightarrow{e}^{R}_{ij} = \frac{\overrightarrow{r}_j - \overrightarrow{r}_i}{\left|\left|\overrightarrow{r}_j - \overrightarrow{r}_i\right|\right|}$,  $\overrightarrow{e}^{T}_{ij} = \frac{\overrightarrow{e}^z\times\overrightarrow{e}^R}{\left|\left|\overrightarrow{e}^z\times\overrightarrow{e}^R\right|\right|}$, 
$\overrightarrow{e}^{S}_{ij} = \frac{\overrightarrow{e}^T\times\overrightarrow{e}^R}{\left|\left|\overrightarrow{e}^T\times\overrightarrow{e}^R\right|\right|}$.

The continuity equation (\ref{SPH-continuity}) written in terms of the volume strain rate has an analytical solution (where $\dot{\varepsilon_i}$ is assumed constant within the integration step), which allows us to apply the following expression~\cite{Parshikov:JCP:2002} to calculate the density at each integration step:
\begin{equation}\label{eq:time_integration}
	\hat{\rho_i} = \rho_i\exp\left(-\tau\dot{\varepsilon_i}\right),
\end{equation}
where $\tau$ is the time step. The integration of (\ref{SPH-momentum}) and (\ref{SPH-energy}) is performed utilising the Euler scheme.

It is worth explaining the form of the continuity equation (\ref{SPH-continuity}). The equation (\ref{SPH-continuity}) is obtained based on a Taylor series expansion.  For this purpose, let us consider the following sum:
$$\sum\limits_j\frac{m_j}{\rho_j}\overline{U}_{ij}^{\alpha}\frac{\partial W_{ij}}{\partial r_i^{\alpha}},$$
where $\overrightarrow{\overline{U}}_{ij}$ is the velocity at the point midway between the centers of particles $i$ and $j$, $\alpha = x, y, z$, $r^x,\, r^y,\, r^z = x,\, y,\, z$. By the repeated Greek indices we will assume summation. By performing a Taylor series expansion in the neighborhood of the point $\overrightarrow{r}_i$ it is easy to obtain:
\begin{equation}\label{continuity-expansion}
	2\sum\limits_j\frac{m_j}{\rho_j}\left(\overline{U}_{ij}^{\alpha}-U_i^{\alpha}\right)\frac{\partial W_{ij}}{\partial r_i^{\alpha}} =  \sum\limits_j\frac{m_j}{\rho_j}\frac{\partial U^{\alpha}_i}{\partial r^{\beta}}\left(r_j^{\beta}-r_i^{\beta}\right)\frac{\partial W_{ij}}{\partial r_i^{\alpha}} + O(h).
\end{equation}
The expression (\ref{continuity-expansion}) can be rewritten in the following form:
\begin{equation}\label{continuity-expansion-2}
	2\sum\limits_j\frac{m_j}{\rho_j}\left(\overrightarrow{\overline{U}}_{ij}-\overrightarrow{U}_i\right)\cdot\nabla_iW_{ij} =  \mathbf{L}_i:\left(\nabla\otimes\overrightarrow{U}_i\right) + O(h).
\end{equation}
If we substitute $\nabla_iW_{ij}\rightarrow \mathbf{L}_i^{-1}\cdot \nabla_iW_{ij}$ in equation (\ref{continuity-expansion-2}), we get a different equation that more accurately approximates the velocity divergence:
\begin{equation}\label{continuity-expansion-3}
	2\sum\limits_j\frac{m_j}{\rho_j}\left(\overrightarrow{\overline{U}}_{ij}-\overrightarrow{U}_i\right)\cdot\left(\mathbf{L}_i^{-1}\cdot \nabla_iW_{ij}\right) =  \mathbf{I}:\left(\nabla\otimes\overrightarrow{U}_i\right) + O(h) = \nabla\cdot\overrightarrow{U}_i + O(h),
\end{equation}
where $\mathbf{I}$ is an identity matrix. Taking  the contact velocity $\overrightarrow{U_{ij}^{*}}$ as an estimate of the velocity $\overrightarrow{\overline{U}}_{ij}$, we obtain equation (\ref{SPH-continuity}). The equations (\ref{SPH-momentum}) and (\ref{SPH-energy}) can also be derived from the Taylor series expansion assuming a regular packing of SPH particles (e.g., cubic). 

The use of the correction near the free surface violates the boundary conditions. Therefore, $\mathbf{L}_i = \mathbf{I}$ is set for particles near the free surface. The algorithm for determining the boundary particles was taken from the work of \cite{Marrone2010}. The determination of the boundary particles is based on using the minimum eigenvalue $\lambda_{min}$ of the matrix $\mathbf{L}_i^{-1}$. If $\lambda_{min} < 0.75$ then the particle is considered as a boundary particle. In this paper, a particle is considered to be close to the boundary if there is a boundary particle within its smoothing length.

To avoid problems related to singularity or poor conditioning of the matrix $\mathbf{L}_i$, the following technique is used: when $\det(\mathbf{L}_i) < \varepsilon$, where the parameter $\varepsilon$ is chosen depending on the problem and usually has values $0.01 - 0.1$, then $\mathbf{L}_i = \mathbf{I}$ is set.

The strain rate tensor is calculated by the formulas:
\begin{equation*}
	\frac{d\hat{\varepsilon}_i}{dt} = \frac{1}{2}\left(\nabla\otimes\overrightarrow{U_i} + \left(\nabla\otimes\overrightarrow{U_i}\right)^\text{T} \right),
\end{equation*}
where
\begin{equation}\label{eq:velocity-derivative-sph}
\nabla\otimes\overrightarrow{U_i} = \sum\limits_j\frac{m_j}{\rho_j}2( \mathbf{L}_i^{-1}\cdot\nabla_iW_{ij})\otimes \left(\overrightarrow{U}^{*}_{ij} - \overrightarrow{U_i}\right).
\end{equation}

Then the numerical scheme for the rate of change of the deviatoric stress tensor (\ref{eq:dSdt}) will be
\begin{equation}\label{eq:dSdt1}
	\frac{d\hat{S}_i}{dt}= 2G\left(\frac{d\hat{\varepsilon}_i}{dt} - \frac{1}{3}\text{tr}\left(\frac{d\hat{\varepsilon}_i}{dt}\right)\hat{I}\right).
\end{equation}

The rotational speed is determined as follows:
\begin{equation}\label{eq:rotation}
\overrightarrow{\omega}_i = \nabla\times\overrightarrow{U_i},
\end{equation}
where the velocity derivatives can be obtained by combining the derivatives from (\ref{eq:velocity-derivative-sph}). Then the correction for the rotation of the stress tensor per integration step is:
\begin{equation}\label{eq:timeDeviator}
\hat{S}_i(t+\tau) = \hat{\Omega}_i(\hat{S}_i(t) +\dot{\hat{S}}_i \tau) \hat{\Omega}^T_i.
\end{equation}
Assuming that the rotation angle is small within the integration step, let us write the antisymmetric matrix $\hat{\Omega}_i$ in index form: 
\begin{equation*}
	\Omega^{kl}_i = \delta^{kl}+\varepsilon^{klm} \omega_i^m \tau,
\end{equation*}
where $\varepsilon^{klm}$ is the Levy--Civita symbol and $\delta^{kl}$ is the unit tensor.

\subsection{Validation of the modelling method}
\subsubsection{Elastic-plastic Riemann problem}
Let us consider the Riemann problem in an elastic-plastic material with the Wilkins~\cite{Alder1964} model. The equation of state is given in the form of a logarithmic dependence
\begin{equation}\label{eq:log_eos}
	P = \rho_0 c_a^2  \ln(V_0/V) = B\ln(V_0/V),
\end{equation}
where $V = 1/\rho$.
The initial states to the left and to the right of the discontinuity (initial conditions) are given in the following form
\begin{equation}\label{eq:initial-condition-elpl}
	(\sigma_{xx}, P, \rho, u^x)=
	\begin{cases}
		(\sigma_1, P_1,  \rho_1,  0) & \text{if } x<0\\
		(0, 0, \rho_0, 0) & \text{if } x\geq 0
	\end{cases},
\end{equation}
where $P_1 = - \sigma_1 - 2/3 Y$, $\rho_1 = \rho_0\exp(P_1/B)$. As a material, consider aluminum with $Y = 0.3\,$GPa, $B = 73\,$GPa, $G = 23\,$GPa, $\rho_0 = 2688.9\,$kg/m$^3$. The shear modulus $G$ is assumed to be constant. $\sigma_1 = -4\,$GPa.

\begin{figure}
	\begin{center}
		\includegraphics[width=\linewidth]{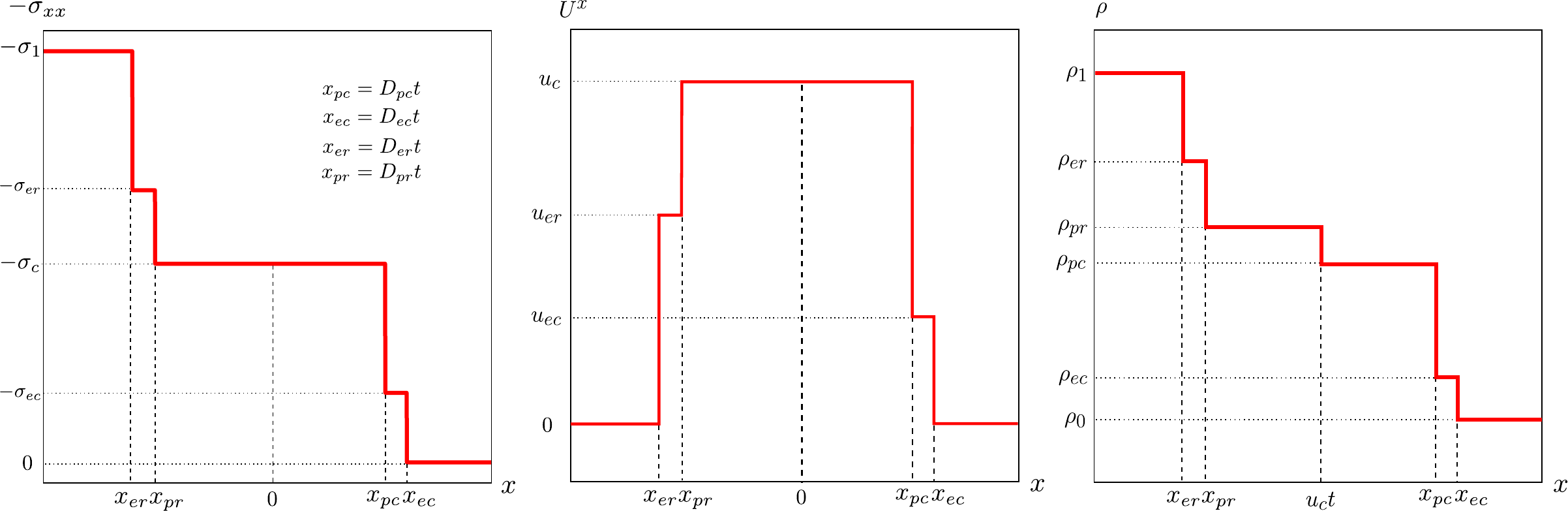}
	\end{center}
	\caption{\label{elastic-plastic-variables} Wave configuration in the elastic-plastic Riemann problem}
\end{figure}

The symbols used below are shown in the Fig.~\ref{elastic-plastic-variables}.

In the flat symmetric case $S_{xx} = -2S_{yy}=-2S_{zz}$, so $S_{xx}$ is bounded by modulo $2Y /3$.
And the relation (\ref{eq:dSdt}) is transformed into a differential relation, which can be integrated along the particle trajectory:
\begin{equation}\label{eq:Hook}
dS_{xx} = \frac{4G}{3}\frac{dV}{V}.
\end{equation} 

Now we give a method for obtaining an analytical solution. We will consider a configuration in which two waves run to the left and to the right: elastic and plastic. The elastic and plastic compression waves run to the right, and the elastic and plastic rarefaction waves run to the left. The elastic rarefaction wave has twice the amplitude, as the deviator in the rarefaction wave changes from $-2Y/3$ in the initial plastic compressed state to $+2Y/3$ in the plastic stretched state. The elastic wave is followed by a plastic wave, where the substance experiences only volumetric deformations.
We start the solution construction with elastic waves, from which we then pass to plastic waves, taking into account the condition of equality of stresses and velocities at the contact: 
\begin{equation}\label{eq:contact}
\sigma_{1c}=\sigma_{0c} = \sigma_c,\;\;\; u_{1c}=u_{0c} = u_c. 
\end{equation}
For simplicity of notation, the indices ``$xx$'', ``$x$'' are omitted since the problem is assumed to be one-dimensional.

When constructing the exact solution (see, e.g., \cite{Menshov2013} for details), we will use the law of conservation of mass
\begin{equation}\label{eq:mass1}
	\rho_l (u_l-D) = \rho_r (u_r-D) = a,
\end{equation}
 and the Michelson-Reley line
\begin{equation}\label{eq:Michelson}
	\sigma_l  - \sigma_r = a^2(V_l - V_r),
\end{equation}
where $D$ is the velocity of the wave (discontinuity velocity).

Since the pressure depends only on the density, the law of conservation of energy will not be needed to construct the solution. 

The evolution of deviatoric stresses in an elastic precursor can be calculated from Hooke's law (\ref{eq:Hook}) and the yield stress condition:
\begin{equation}\label{1}
S_{ec} = -\frac{2}{3} Y = -\frac{4}{3}G\ln\left(\frac{V_0}{V_{ec}}\right).
\end{equation}
By employing equation (\ref{1}), the density and stress in the elastic precursor can be calculated:
\begin{equation}\label{eq:V_el}
	V_{ec} = V_0 \exp(-\frac{Y}{2G}),
\end{equation}
	$$\sigma_{ec} = S_{ec} - P(V_{ec}) = -\frac{2Y}{3}-\frac{Y B}{2 G}.$$

The velocity of an elastic precursor can be expressed from the law of conservation of mass (\ref{eq:mass1}):
\begin{equation}\label{D_ec}
	D_{ec} = \frac{\rho_{ec}u_{ec}}{\rho_{ec} - \rho_0}.
\end{equation}
Using the relation (\ref{eq:Michelson}) taking into account (\ref{D_ec}) and the fact that $u_0 = 0$ and $\sigma_0 = 0$ we obtain the equation to determine the velocity in the elastic precursor:
$$\sigma_{ec} = \rho_0^2\left(\frac{\rho_{ec}u_{ec}}{\rho_{ec} - \rho_0}\right)^2(V_{ec} - V_0),$$
from where we find 
$$u_{ec} = \sqrt{\sigma_{ec}(V_{ec} - V_0)}.$$

Let us express the elastic precursor velocity $D_{ec}$ through the input data:
$$D_{ec} = \sqrt{\frac{(2/3+B/(2G))Y}{1-\text{exp}\{-Y/(2G)\}}}.$$
Note that if we decompose the exponent in the denominator near zero to a linear term, we obtain that the precursor velocity is equal to the longitudinal sound speed $c_l(V_0)$ given by (\ref{eq:cl}).

The elastic rarefaction wave can be calculated after determining the density of a compressed material with a given stress $\sigma_1$ from the initial condition (\ref{eq:initial-condition-elpl}) and the condition that the material is at the elastic-plastic limit in compression. The change in the deviatoric part of the stresses in an elastic rarefaction wave is twice as large as in an elastic precursor and, therefore, the material density is calculated after a similar application of the equation for the change in deviatoric stress (\ref{eq:Hook}):
\begin{equation*}
\frac{4}{3}Y = \frac{4}{3}G \ln\left(\frac{V_{er}}{V_1}\right),
\end{equation*}
and therefore
\begin{equation}\label{eq:V_er}
V_{er} = V_1 \text{exp}(Y/G).
\end{equation}
The stress in the elastic rarefaction wave can be calculated from the initial condition and the stress deviator and pressure jump:
$$\sigma_{er} = \sigma_1 + \frac{4}{3}Y - B\ln\left(\frac{V_0}{V_{er}}\right) + B\ln\left(\frac{V_0}{V_1}\right) = \sigma_1 + \frac{4}{3}Y + B\ln\left(\frac{V_{er}}{V_1}\right) = \sigma_1 + \frac{4}{3}Y + \frac{BY}{G}.$$

The velocity of matter in an elastic wave of rarefaction and the velocity of propagation of this wave are defined in the same way as in the case of an elastic wave of compression (elastic precursor), so let us immediately write out the final equations:
$$u_{er} = \sqrt{(\sigma_{er}-\sigma_1)(V_{er} - V_1)},$$
$$D_{er} = \frac{\rho_{er}u_{er}}{\rho_{er} - \rho_1}.$$
It should be noted that similarly to the compression wave, when expanded in a Taylor series taking into account $Y\ll G$ to the linear term, we obtain $D_{er}=c_l(V_1)$. 

The elastic waves in compression and rarefaction are followed by plastic waves, which have velocities $D_{pc}$ and $D_{pr}$. The stress variation is only due to the spherical part of the stresses, since the deviatoric components are confined to the Mises cylinder.

A compressive plastic wave is characterized by a step change in stress, which can be calculated from the equation of state (\ref{eq:log_eos}):
\begin{equation}\label{eq:delta_sigma_pc}
\Delta \sigma_{pc} = -B \ln(V_{ec}/V_{pc}),
\end{equation}
where $V_{ec}$ is defined by (\ref{eq:V_el}).
On the other hand, the stress jump is defined by equation (\ref{eq:Michelson}). Expressing  $a_{pc}$ from the law of conservation of mass we obtain
$$a_{pc} = \frac{u_c - u_{ec}}{V_{pc} - V_{ec}},$$ 
substituting the obtained expression into (\ref{eq:Michelson}) and using (\ref{eq:delta_sigma_pc}) we obtain the following equation:
\begin{equation}\label{1st-equation-of-system}
	B\ln\left(\frac{V_{pc}}{V_{ec}}\right) = \frac{(u_c - u_{ec})^2}{V_{pc} - V_{ec}}.
\end{equation}

Similarly, we obtain the equation on the plastic rarefaction wave:
\begin{equation}\label{2nd-equation-of-system}
	\Delta\sigma_{pr} = \ln\left(\frac{V_{pr}}{V_{er}}\right) = \frac{(u_c - u_{er})^2}{V_{pr} - V_{er}}.
\end{equation}
 
 Using the condition of equality of stresses on the contact (\ref{eq:contact}) we obtain another equation:
 \begin{equation*}
 	\sigma_{ec} + \Delta\sigma_{pr} = \sigma_c = \sigma_{er} + \Delta\sigma_{pc}.
 \end{equation*}
 Or, which is the same thing,
  \begin{equation}\label{3rd-equation-of-system}
	\sigma_{ec} + B\ln\left(\frac{V_{pc}}{V_{ec}}\right) = \sigma_{er} + B\ln\left(\frac{V_{pr}}{V_{er}}\right).
  \end{equation}

The resulting equations (\ref{1st-equation-of-system})-(\ref{3rd-equation-of-system}) form a system that is solved numerically with respect to $V_{pr},$ $V_{pc}$ and $u_c$.
The velocities of compressional and rarefaction plastic waves can be expressed from (\ref{eq:mass1}):
$$D_{pc} = \left(\frac{V_{pc}}{V_{ec}}u_{ec}-u_c\right)/\left(\frac{V_{pc}}{V_{ec}}-1\right),$$
$$D_{pr} = \left(\frac{V_{er}}{V_{pr}}u_c-u_{er}\right)/\left(\frac{V_{er}}{V_{pr}}-1\right).$$

The contact (\ref{eq:contact}) was at rest at the initial moment of time. According to the law of conservation of mass, the flow of matter on the left at each moment of time is equal with opposite sign to the flow on the right, and the contact surface moves with constant velocity $u_c$. 

\begin{figure}
 	\begin{center}
 		\includegraphics[width=\linewidth]{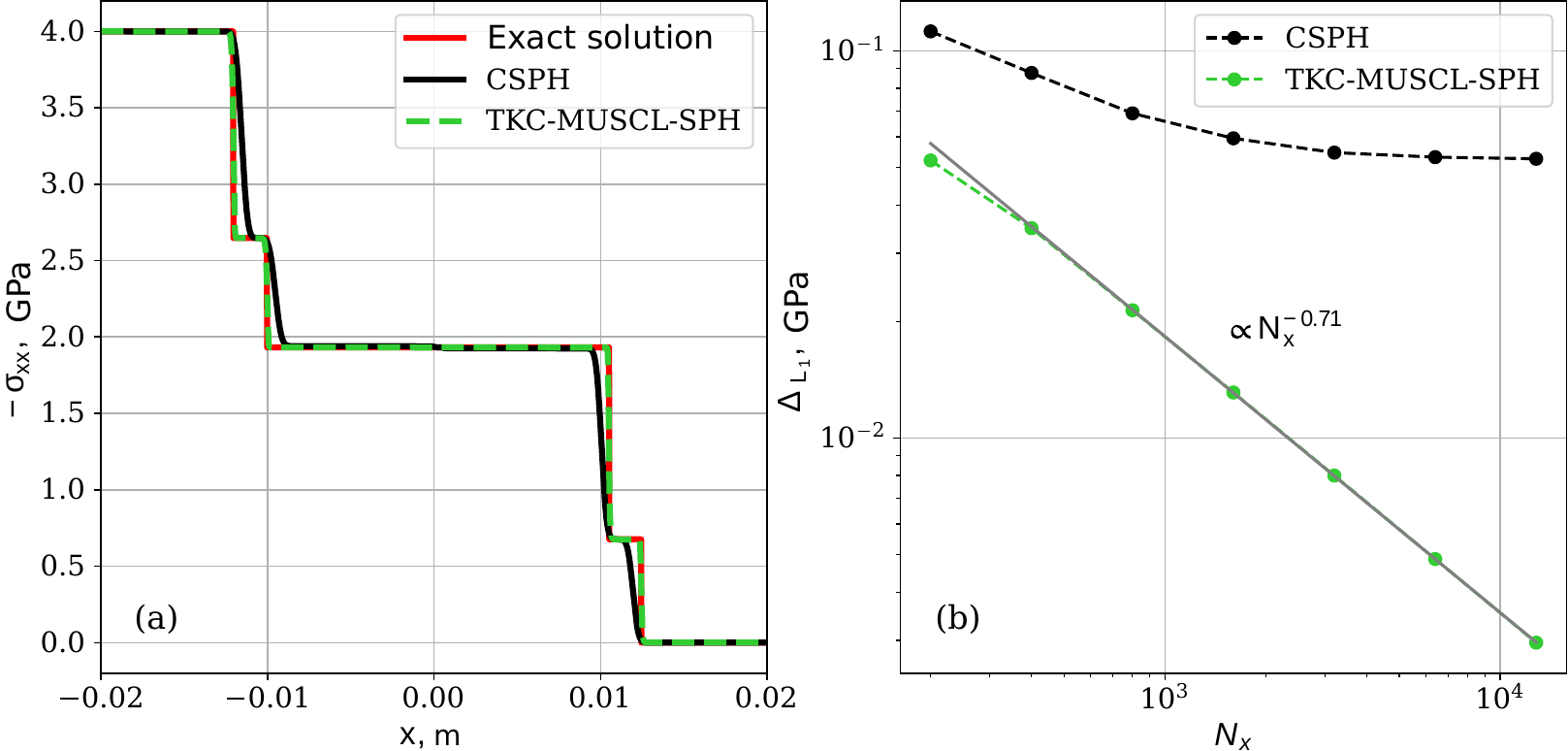}
 	\end{center}
 	\caption{\label{elastic-plastic-convergence} (a) Stress profiles in the elastic-plastic rupture decay problem in copper obtained by the standard CSPH \cite{Parshikov:JCP:2002} and the proposed TKC-MUSCL-SPH method. (b) Dependence of the $L_1$ norm error in $\sigma_{xx}$ stress component on the number of particles along the $x$-axis for the standard CSPH \cite{Parshikov:JCP:2002} and the proposed TKC-MUSCL-SPH method. $\theta = 0.5.$}
 \end{figure}

Using the obtained analytical solution (\ref{eq:initial-condition-elpl}), we validate the proposed TKC-MUSCL-SPH method. The length of the computational domain along the $x$-axis was $0.05\,$m. The modeling was continued until the time $2\,\mu$s. The problem was considered in 2D formulation with a smoothing kernel Wendland C$^2$ and smoothing length parameter $\theta = 0.5$. Calculations with different numbers of particles along the $x$-axis were performed to investigate convergence. Figure \ref{elastic-plastic-convergence}(a) shows the profiles of the stress tensor component $\sigma_{xx}$ obtained by the CSPH method and the proposed TKC-MUSCL-SPH method, as well as the exact solution. It can be seen that the proposed method has significantly lower numerical diffusion compared to the CSPH method. In addition, the kernel correction allows us to obtain wave propagation velocities closer to the analytical ones. Figure \ref{elastic-plastic-convergence}(b) shows the dependence of the stress error in $L_1$ norm
 $$\Delta_{L_1} = \frac{1}{N}\sum\limits_{i = 1}^{N}\left|\sigma^{xx}_{\text{SPH}}(x_i) - \sigma^{xx}_{\text{exact}}(x_i)\right|$$
on the number of particles along the $x$-axis. It is well seen that the accuracy of the proposed TKC-MUSCL-SPH method is much higher than that of the CSPH method.

 \subsubsection{Longitudinal sound wave}
Consider the problem of longitudinal sound wave propagation in copper.
The initial density of copper is $\rho_0 = 8.96$~g/cm$^3$. The equation of state is given in the form:
$$p = c_0^2(\rho-\rho_0),$$
where $c_0 = 4.148\,$km/s. The shear modulus $G$ is calculated assuming a fixed Poisson's ratio $\nu = (3B-2G)/(2(3B+G)) = 0.33$.
Initially, the following velocity field is given:
$$U^x = 5\exp(-2x^2).$$
Further, the original perturbation is decomposed into two waves running in opposite directions with amplitudes half as small as the original one (i.e., 2.5 m/s each). Comparison of the obtained result is made with the reference solution obtained by the mesh-based Lagrangian method~\cite{Samarsky1992}.

 \begin{figure}
	\begin{center}
		\includegraphics[width=\linewidth]{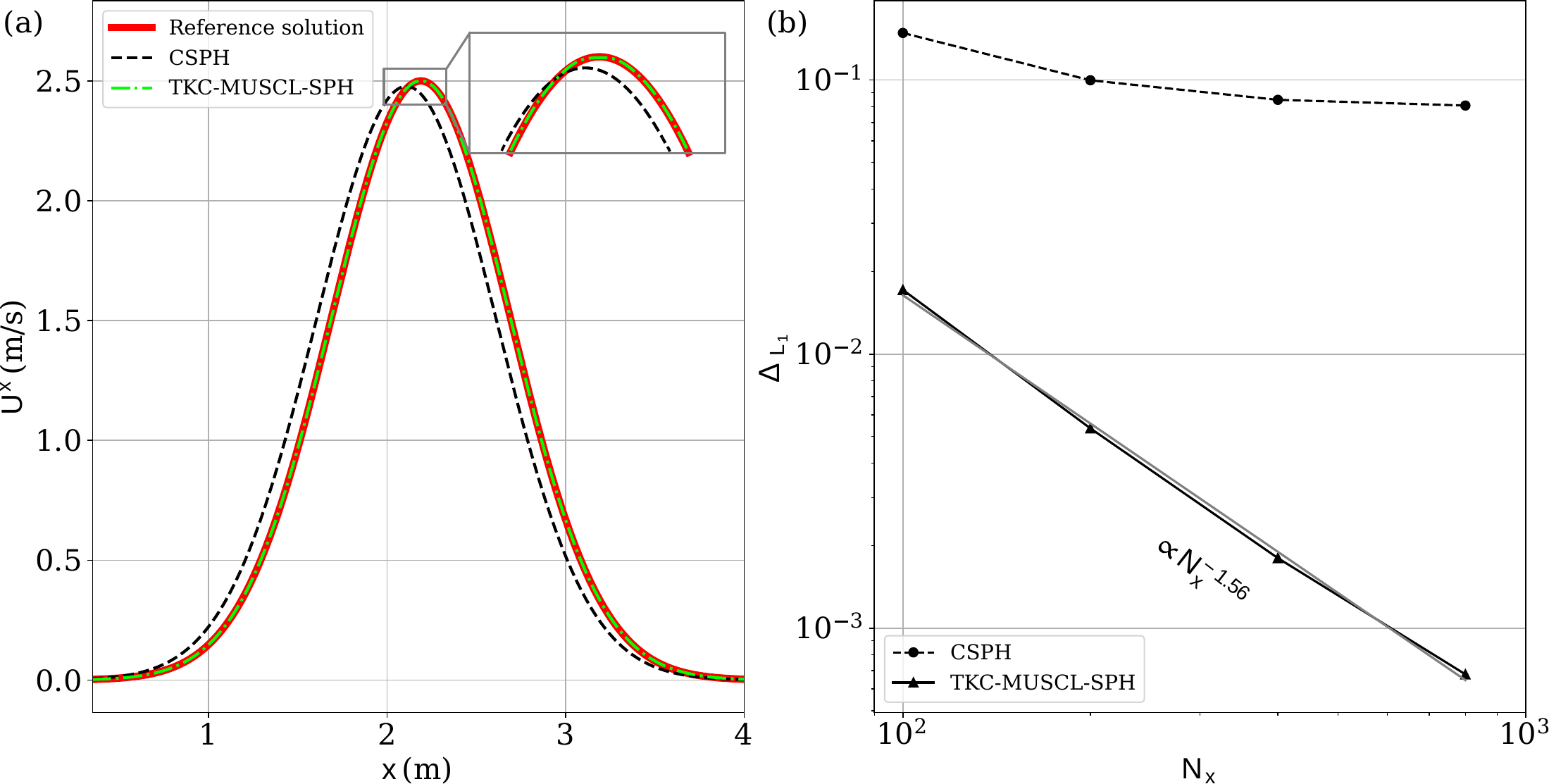}
	\end{center}
	\caption{\label{sound-wave-convergence} (a) Wave (running to the right) in copper. Results of modeling by the standard CSPH method \cite{Parshikov:JCP:2002} and the proposed TKC-MUSCL-SPH method. (b) Dependence of the $L_1$ norm error in the $U_{x}$ velocity on the number of particles along the $x$-axis for the standard CSPH \cite{Parshikov:JCP:2002} and the proposed TKC-MUSCL-SPH method. $\theta = 0.6.$}
\end{figure}
Figure \ref{sound-wave-convergence} shows the simulation results obtained by CSPH and TKC-MUSCL-SPH methods of sound wave propagation in copper.   Figure \ref{sound-wave-convergence}(a) shows that the wave centers in the CSPH method are displaced relative to the reference solution, which does not occur in the TKC-MUSCL-SPH simulation.  The $L_1$-norm of the error is calculated as:
$$\Delta_{L_1} = \frac{\sum\limits_b\left| U^x_b - \left(U^x_b\right)^{\mathrm{reference}} \right|}{N}.$$

 \subsubsection{Cylindrical Verny problem}
 
 This problem is of interest from the point of view of the accuracy of modeling elastic and plastic waves in the presence of free boundaries. In this problem, the radial motion of a cylindrical shell to the axis of symmetry is modeled until it comes to a complete stop. 
 
 The shell material is modeled using the equation of state in the form:
 $$p = c_0^2(\rho - \rho_0) + (\Gamma-1)\rho e.$$	
 	
 The initial outer and inner radii of the shell are $R_1 = 0.1\,$m, $R_0 = 0.08\,$m. The material of the shell is aluminum with density $\rho = 2785\,$kg/m$^3$, shear modulus $G = 27.6\,$GPa, yield strength $Y = 0.3\,$GPa, bulk speed of sound $c_0 = 5328\,$m/s, $\Gamma = 2$. 
 
 The initial velocity distribution is
 $$u(r) = U_0\frac{R_0}{r},$$
 where $U_0 = 208.55\,$m/s.
 
 The velocity $U_0$ required to reach the inner radius at stop $r_0'$ is determined by the formula (see, e.g., \cite{Howell2002})
  $$U_0 = \sqrt{\frac{2YF(\alpha,\lambda)}{\sqrt{3}\rho\ln(R_1/R_0)}},$$
  where 
  $$F(\alpha, \lambda) = \int\limits_{\lambda}^1x\ln\left(1+\frac{2\alpha + \alpha^2}{x^2}\right)dx,$$
  $$\alpha = \frac{R_1-R_0}{R_0},\,\,\lambda=\frac{r_0'}{R_0}.$$
From here it is not difficult to obtain that at the given parameters the inner radius at stopping is $r_0' = 0.0667\,$m.

 \begin{figure}
	\begin{center}
		\includegraphics[width=\linewidth]{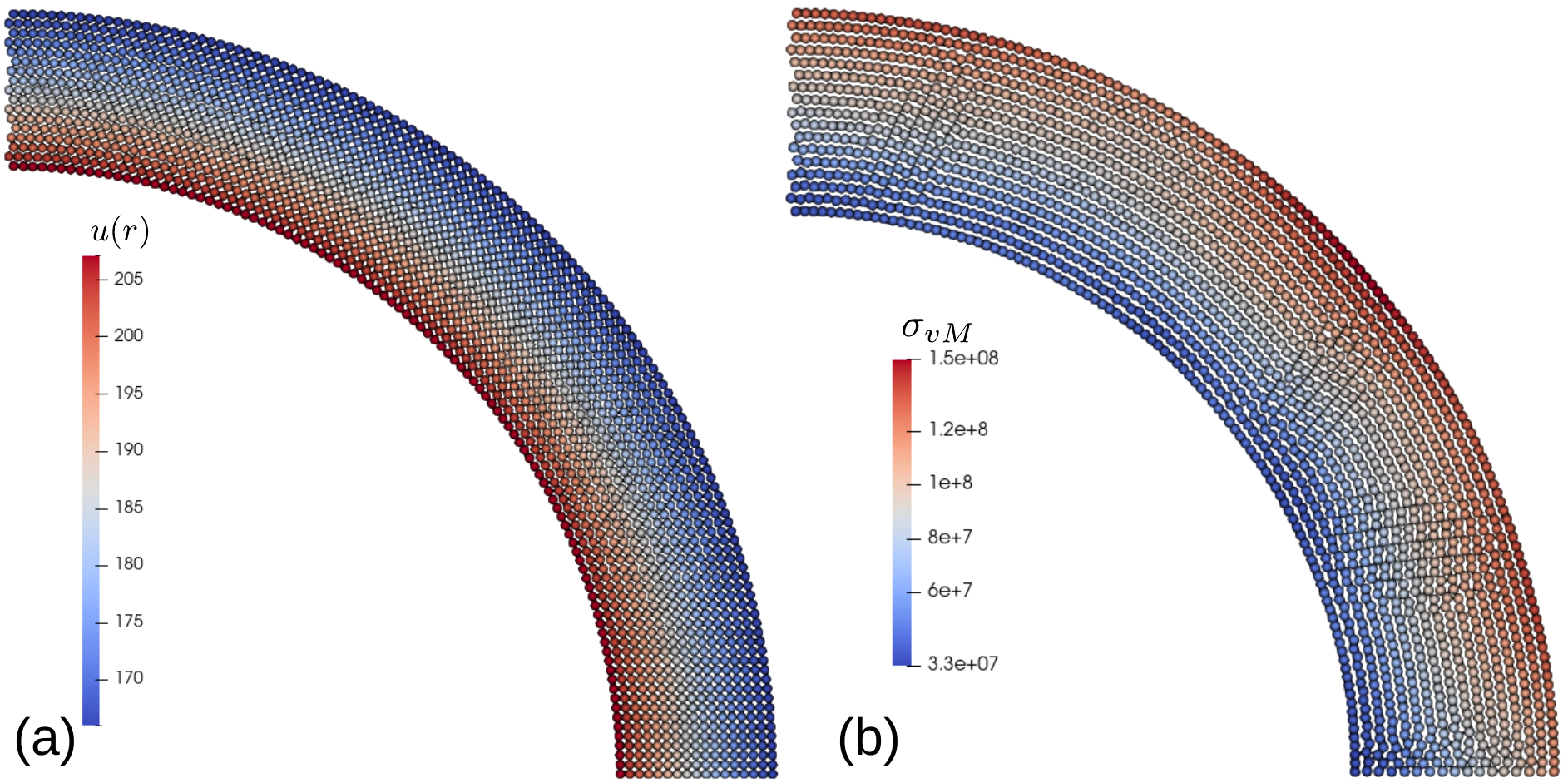}
	\end{center}
	\caption{\label{verni-packing} Initial (a) and final (b) packing of SPH particles in the cylindrical Verny problem.}
\end{figure}
\begin{figure}
	\begin{center}
		\includegraphics[width=0.8\linewidth]{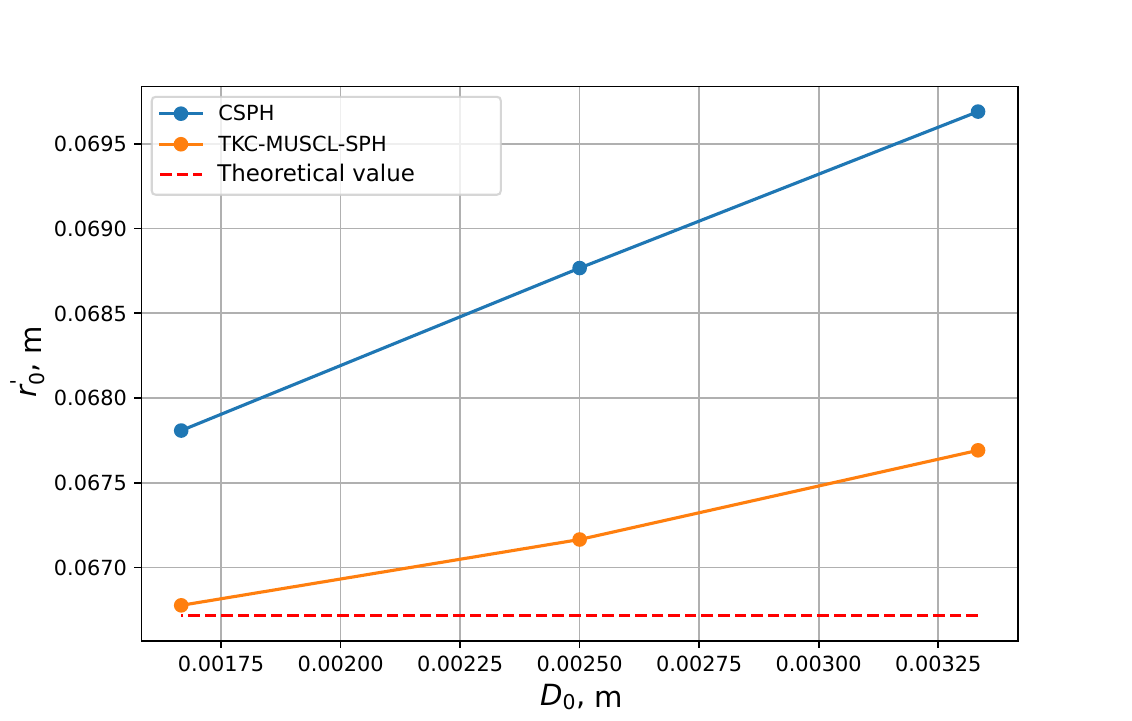}
	\end{center}
	\caption{\label{Internal-stoppnig-radius} Dependence of the internal radius at compression stopping $r_0'$ on the initial particle size $D_0$ for the standard CSPH method \cite{Parshikov:JCP:2002} and the proposed TKC-MUSCL-SPH method at $\theta = 0.5.$}
\end{figure}

 
 Figure \ref{verni-packing}(a) shows the initial packing of SPH particles and the radial velocity. Figure \ref{verni-packing}(b) shows the final packing of SPH particles in the cylindrical Verni problem and the equivalent stress distribution $\sigma_{vM} = \sqrt{(S_{xx}^2+S_{yy}^2)-S_{yy}S_{xx}+3S_{xy}^2}$. Figure \ref{Internal-stoppnig-radius} shows the internal radii at stopping obtained by CSPH and TKC-MUSCL-SPH methods. The values show that the proposed TKC-MUSCL-SPH method is more accurate than the original CSPH method.

\section{Moving window technique}
\begin{figure}[h]
	\centering
	\includegraphics[width=1\columnwidth]{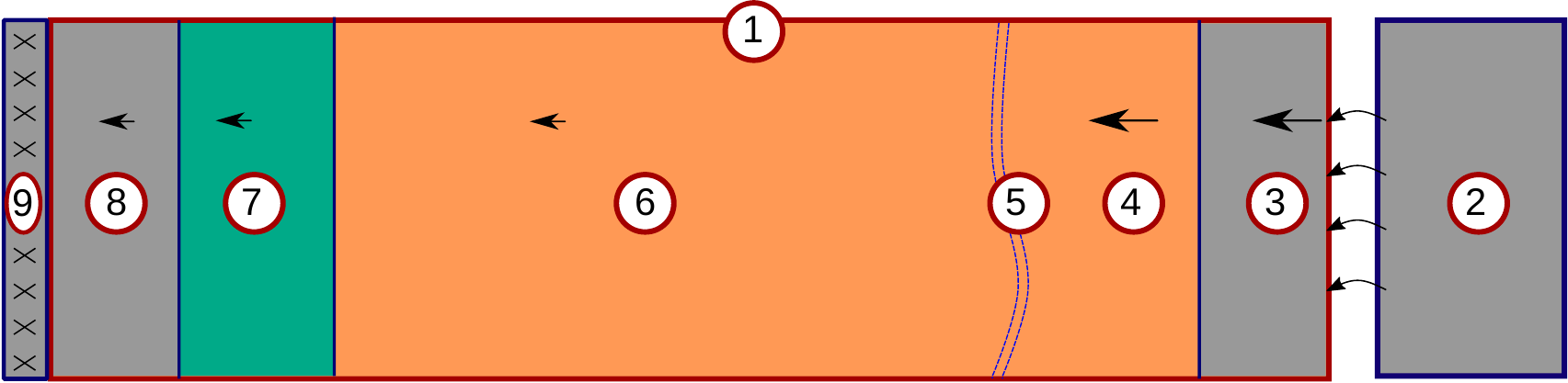}
	\caption{1 --- moving observation window; 2 --- sample of the medium prepared for particle entry into the observation window; 3 --- particle entry region; 4 --- flow zone through the shock wave front; 5 --- shock wave front; 6 --- relaxation zone; 7 --- particle exit region 1; 8 --- particle exit region 2; 9 --- particle removal from the observation window.}
	\label{fig:scheme}
\end{figure}
The computational domain has the shape of a rectangle (when modeling 3D flows, it would have the shape of a rectangular parallelepiped).
Figure~\ref{fig:scheme} shows a schematic of observation window 1 for the 2D case. SPH particles enter the observation window 3 through the boundary on the right and exit into the material flow zone ahead of the shock front through the boundary on the left. 

The particle exit boundary from the observation window has a given coordinate $x = x_l$, crossing which the SPH-particles are removed from the calculation. Inside the observation window there are two regions adjacent to the exit boundary. Particles with coordinates $d_{out}^2 < x-x_l < d_{out}^1$ and $x-x_l < d_{out}^2$ form two exit boundary regions, where $d_{out}^i$ is the width of regions $i=1,2$. The particles enter the exit region 7 from the relaxation zone 6 behind the shock front. The interaction of particles from the exit region 7 with particles from the relaxation zone leads to an exchange of momentum between them and, accordingly, to the formation of a shock wave in the material. In this case, the state of interacting SPH-particles changes according to the solution of the system of SPH-equations representing the conservation laws. 

In the output region 8, the particle motion occurs with a constant velocity. The state of particles~\cite{HAFTU2022108377} is also fixed. Fixation of the particle state is necessary to exclude the influence of the free surface of the medium, which is beyond the left boundary of the observation window. 

Particles leaving the observation window are assigned a constant exit velocity $u_x = u_{out}$, and their transverse velocity component is zero. All other parameters (density, energy and other mechanical and thermodynamic quantities) change according to the interaction of the SPH-particle with its environment. Note that by setting the exit velocity of the particles relative to the entry velocity, we determine the velocity of the SW $u_s$ and the mass velocity behind it $u_p$ in the stationary coordinate system associated with the SW front. 

During the modeling process, the particles are shifted relative to their initial position, and particles from the input region 3 to the flow zone 4 in front of the shock front can arrive at each computational step. The input region 3 itself is replenished only when the list of neighbors for all SPH particles is updated, i.e., every few integration steps ($\approx$ 10 steps). Therefore, inside the input region 3 the particle displacement can be up to several layers of SPH-particles with the formation of an unfilled zone adjacent to the right boundary of the observation window. Algorithmically, the process of filling this empty zone is organized as follows: behind the right border of the observation window a sample of the medium 2 with a thickness of several layers of particles and repeating the structure of the medium is created. This sample transfers every few integration steps a part of its particles to the input region 3 to fill the space vacated to the right of the sample. When the sample to be fed ends, it is cycled, since it is prepared under periodic boundary conditions.

The initial filling with particles of the calculation region in the observation window is organized by periodic repetition along the direction of shock wave propagation $X$ of some basic sample, which is a sample of porous copper, enclosing one period of a prearranged structure. Our calculations are performed in flat 2D geometry, the pores have a circular shape and are located in the nodes of a square lattice. 

The array of particles of the mentioned basic sample, by means of which the structure of the medium in the computational domain is created, is ordered in ascending order of coordinate $X$. This is done in order to simplify the search for the boundary in sample 2, which separates the particles necessary for placement in region 3 from the rest of sample 2. 
The coordinates of the inserted particles are transformed to the coordinates of the reference frame for the computational domain. 

The entry (region 3) and exit (regions 7 and 8) regions shown in the figure~\ref{fig:scheme} and fixing the SPH-particle velocity regions from the boundary conditions have a size along the particle motion direction equal to four initial SPH-particle sizes, which ensures that there is no influence of the sample edges on the particles inside the SPH-modeling region.
\begin{figure}[h]
	\centering
	\includegraphics[width=1\columnwidth]{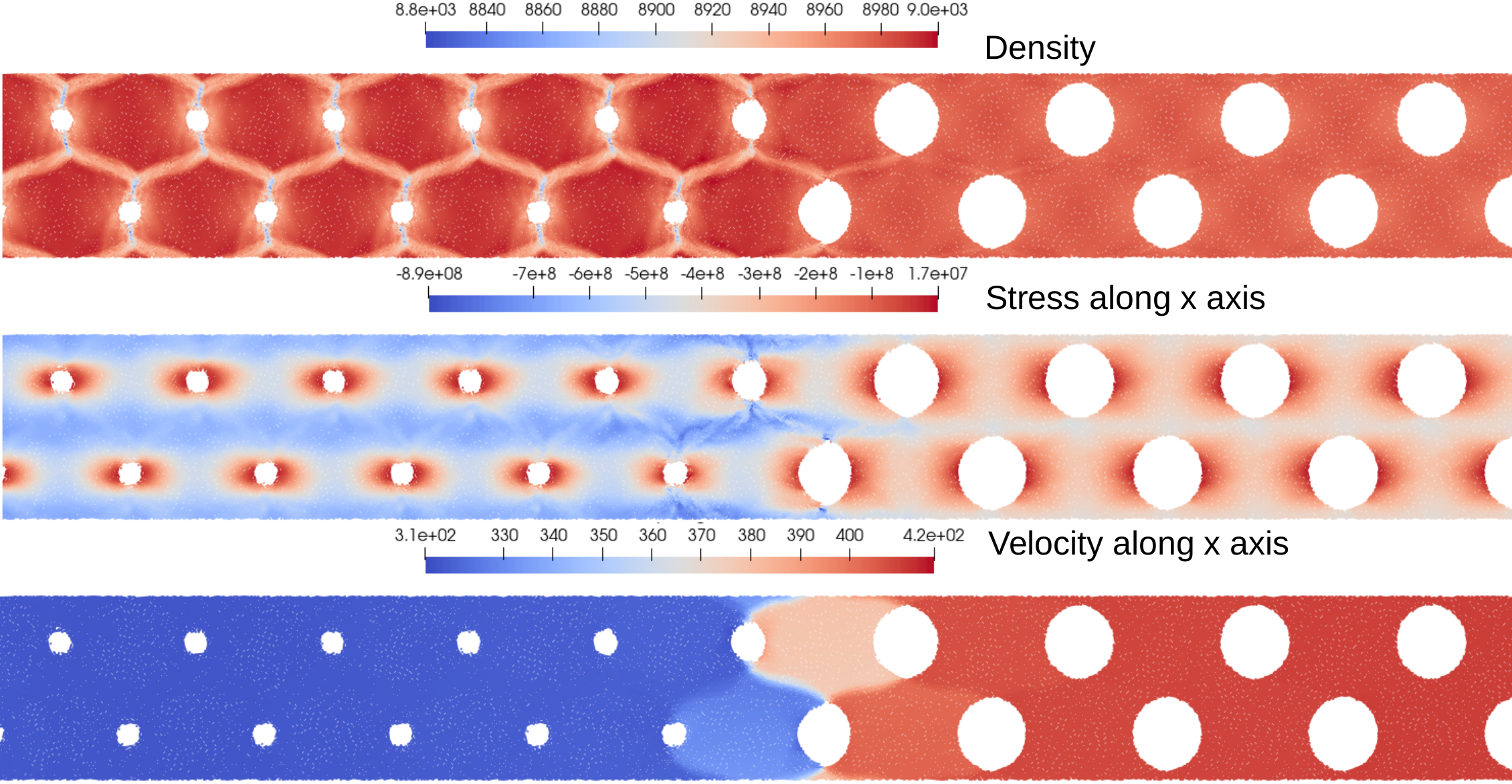}
	\caption{\label{copper-example-100} Density, velocity, and stress tensor component $\sigma^{xx}$ distributions in simulations of shock wave propagation through porous copper at $u_p = 100\,$m/s.}
	
\end{figure}

\begin{figure}[h]
	\centering
	\includegraphics[width=0.8\columnwidth]{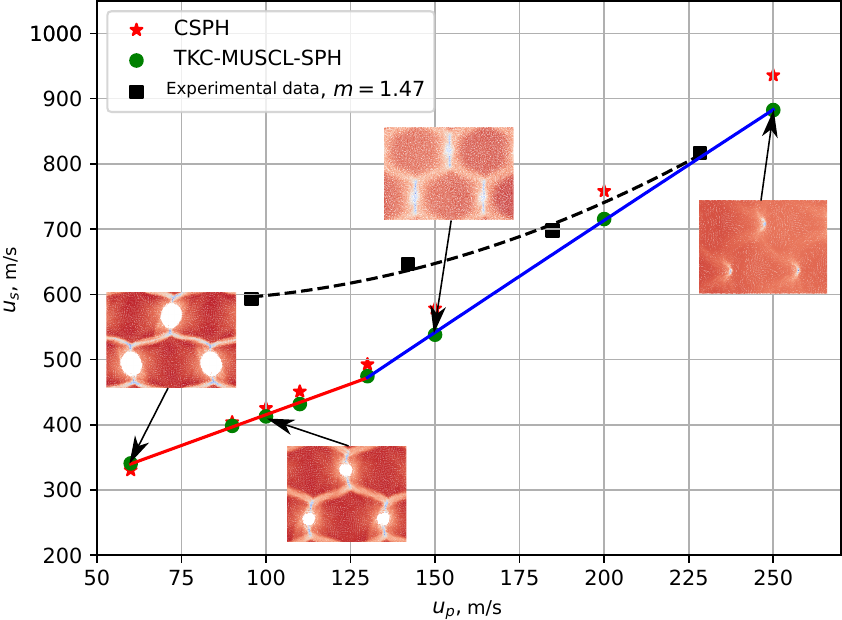}
	\caption{\label{fig:Shock_Hugoniot}   Shock Hugoniot of porous copper at porosity $m = 1.35$ obtained by CSPH and TKC-MUSCL-SPH methods.}
\end{figure}

\subsection{Adjusting the speed of the moving window}
To select the particle exit velocity $u_{out}$ in the problem of modeling a stationary shock wave in porous copper, we define the following value, which characterizes the approximate position of the shock front:
$$\omega = \frac{N - N_0}{\nu N_0},$$
where $N$ is the current number of particles in the computational domain, $N_0$ is the initial number of particles in the computational domain, $\nu$ is a parameter that is chosen so that at the desired position of the shock front in the computational domain $\omega \approx 1$. The value of $\omega$ varies during the calculation, which is not a problem for the $u_{out}$ velocity fitting algorithm, since only the average slope angle $\left<\frac{d\omega}{dt}\right>$ of the $\omega(t)$ function is considered. At the optimal value of $u_{out}$, the average slope angle $\left<\frac{d\omega}{dt}\right>$ should be equal to zero.

For other problems, $\omega$ may have a different form, which is not important for the $u_{out}$ velocity fitting algorithm. The only condition is monotonicity of the average slope angle $\left<\frac{d\omega}{dt}\right>$ of the function $\omega(t)$ depending on the velocity $u_{out}$ in the vicinity of the optimal value.

The algorithm is as follows:
\begin{enumerate}
	\item The velocities $u_p$ and $u_{out}$ and the geometry of the computational domain are set. The simulation of shock wave passage through the porous sample is started.
	\item When the target value $\omega = 1$ is reached, the calculation stops and the obtained state is saved.
	\item The bounds of possible values of velocity $u_{out}$ are set: $u_{out}^{min}$ and $u_{out}^{max}$.
	\item Since, according to the formulated requirements, the signs of the mean slope angle $\left<\frac{d\omega}{dt}\right>$ at $u_{out}^{min}$ and $u_{out}^{max}$ are different, and in the optimal case $\left<\frac{d\omega}{dt}\right> = 0$, the search for the velocity $u_{out}$ is performed by the dichotomy method. For this purpose, at different values of $u_{out}$, the simulation is run from the state saved in step 2, which continues until $n$ elementary material samples are fed into the computational domain (in this work, $n = 3$). The value $\left<\frac{d\omega}{dt}\right>$ is then estimated. The search for the root of the equation $\left<\frac{d\omega}{dt}\right>(u_{out}) = 0$ by the dichotomy method stops when the search range narrows to less than $1\,$m/s.
	\item Since the obtained state might not be stationary, simulations are performed with the found value of $u_{out}$ until the insertion of a sample of length about half of the length of the computational domain occurs. The obtained state is saved.
	\item At the end, the dichotomy algorithm is run again to more accurately determine the velocity $u_{out}$ corresponding to the stationary state.
\end{enumerate}
The velocity fitting algorithm is simplified compared to the work~\cite{Murzov2024}, allowing profiles to be obtained in less simulation time and requiring fewer initial parameters to be tuned. The dichotomy method used in this paper relies only on the monotonicity of $u_s(u_p)$ and the need to choose an initial velocity search interval [$u_{out}^{min}$, $u_{out}^{max}$] such that the values $\left<\frac{d\omega}{dt}\right>$ measured at the ends of the interval have different sign. The second requirement is to estimate $\left<\frac{d\omega}{dt}\right>$ over a sufficiently long time interval $L/u_{in} n$, where $L$ is the size of the underlying feed sample, to eliminate the influence of oscillations in the target measurement due to spatial heterogeneity of the feed sample.

\section{Shock Hugoniot of porous copper}
The equation of state in the form of Mie-Grüneisen and the Mises perfect plasticity model are used as the copper model. The parameters of the model are given in Table \ref{cuprum-model}.

\begin{table*}[t]
	\caption{\label{cuprum-model} Copper model parameters.}
	\centering
	\begin{tabular}{ c|c }
		\hline
		\hline
		Parameter & Value\\
		\hline
		Shear modulus $G,\,$GPa & 43\\
		\hline
		Yield strength $Y,\,$GPa & 0.35\\
		\hline
		Tensile strength $T,\,$GPa & 0.35\\
		\hline
		Grüneisen coefficient $\gamma$ & 1.7\\
		\hline
		$s_a$ & 1.5\\			
		\hline
		$c_a,\,$m/s & 3930\\
		\hline
		$\rho_0,\,$kg/m$^3$ & 8960\\
		\hline
		\hline
	\end{tabular}
\end{table*}

The computational domain is filled with a porous copper sample with porosity $m=\rho/\rho_0$=1.35.
The elastic-plastic SW in porous copper in 2D modeling was calculated for the amplitude where the elastic precursor propagates faster than the plastic wave. 

The sample length along the propagation axis $X$ was approximately 0.56~mm, including the entrance and exit regions. The transverse length $L_y = 20\,\mu$m, and periodic boundary conditions were set at the boundaries. The pore centers are located at the nodes of the square lattice and are circular in shape with a circle radius of $4.06\,\mu$m. A total of 18-30 periods of base specimens were used and copied along the $x$-axis. The base sample has a size of 20$\times$20$\,\mu$m$^2$ and contains one pore lattice period.

Figure~\ref{copper-example-100} shows an example of calculation of SW passage through porous copper at $u_p = 100\,$m/s. Modeling was carried out by the TKC-MUSCL-SPH method.

Figure~\ref{fig:Shock_Hugoniot} shows the shock Hugoniot obtained by modeling the flow of SW through porous copper with porosity $m = 1.35$ using the CSPH (values taken from \cite{Murzov2024}) and TKC-MUSCL-SPH methods.  The obtained shock Hugoniots are close to each other, but the difference between the shock Hugoniots increases with increasing $u_p$. As we have shown in this work, the TKC-MUSCL-SPH method is more accurate than the original CSPH method; therefore, in the authors' opinion, the obtained shock adiabatic is more accurate than the one presented in \cite{Murzov2024}. 

The figure~\ref{fig:Shock_Hugoniot} shows the region of small stress amplitudes in the SW. Two approximating dependencies in the shock Hugoniot have been used. The region with $u_p\leq130$ is approximated by a linear dependence (red line), and the points with $u_p>130$ are approximated by a quadratic dependence (blue line):
\begin{align}
	\nonumber
	u_{s1} &= 226.8 + 1.886 u_p,  && u_p\leq 130, \\
	\nonumber  
	u_{s2} &= 14.58 + 3.573 u_p -0.0003967 u_p^2, && u_p>130.
\end{align}
Note that the presented dependence $u_s(u_p)$ in Fig.~\ref{fig:Shock_Hugoniot} determines the longitudinal stress change $\Delta \sigma_{xx}$ only in the compactification wave, assuming that the density change in the elastic precursor is small
$$\Delta \sigma_{xx} = \rho_0 u_s(u_p) u_p.$$
At the same time, the stress in the elastic precursor $\sigma_{HEL}\approx 0.22$~GPa is a constant value. The final stress behind the wave front is the sum of $\Delta \sigma_{xx}$ and $\sigma_{HEL}$.

The occurrence of a kink in the shock Hugoniot is related to the approach to the yield strength in the amplitude of the compacting wave. The compacting wave completely closes the pores in the simulation with $u_p\geq 150$~m/s.  The two-dimensional distributions in the Fig.~\ref{copper-example-100} for velocity $u_p=100$~m/s show plastic shear bands during the pore compacting process, with the material around the pores plastically deforming, flowing inside the pores, and closing them almost symmetrically, after which the matter in the plastic front between the pores remains in an elastic state. This substance moves as a whole because the speed of sound in copper is an order of magnitude higher than the speed of the plastic compacting wave. Plastic shear bands are formed around such compressed matter. The slippage of the material in the shear bands is also evident from the abrupt change in the tangential velocity component at the slip boundary; this change is on the order of $30$~m/s, as demonstrated in the velocity distribution in the Fig.~\ref{copper-example-100}. As a result, at lower amplitudes of impact, smaller pores remain behind the front of the compacting wave, and the material is in elastic compression, with plastic flow occurring only in the compacting wave. The density of the material behind such a compacting wave remains less than the density of solid copper $\rho_0$, and the average density of SPH-particles is close to the initial $\rho_0$. The latter means that shear stresses in the material behind the compacting wave front keep the material with new pore size, pore spacing ratio, and other structure period in the elastic state at a loading amplitude greater than the initial material can withstand before plastic deformation. 

 \begin{figure}[h]
 	\centering
 	\includegraphics[width=1\columnwidth]{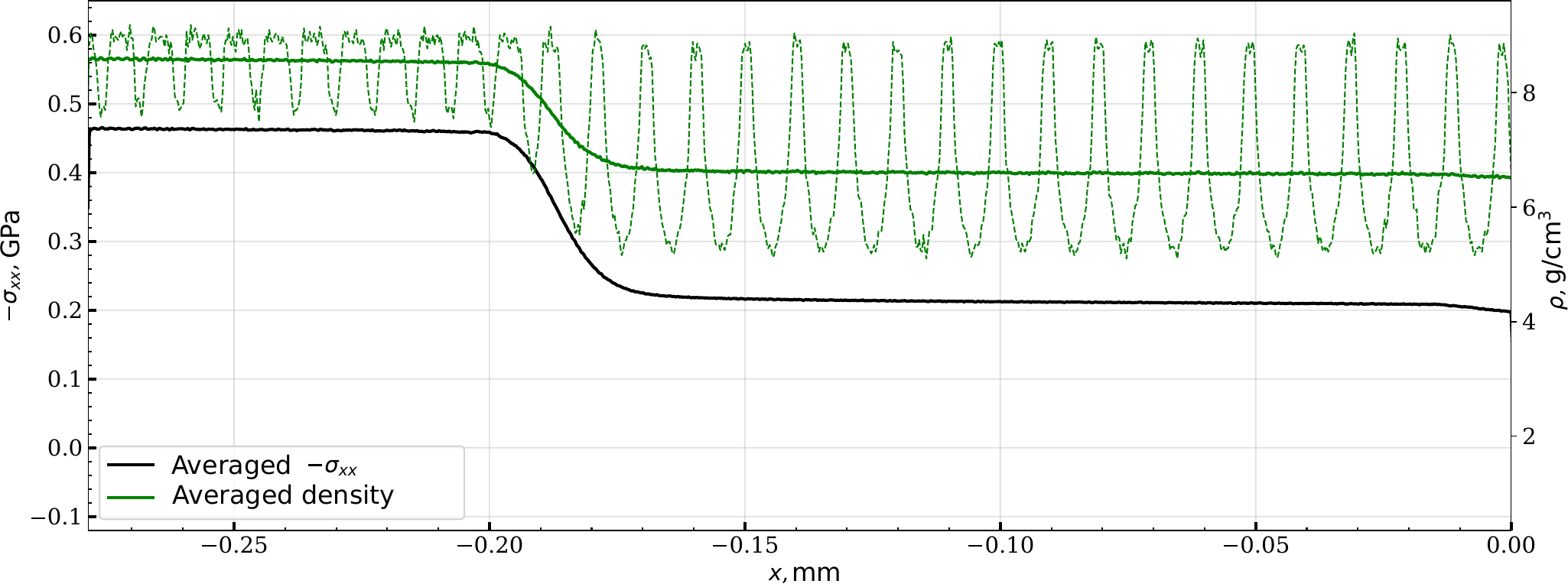}
 	\caption{\label{fig:profiles} Density and longitudinal stress component profiles for a stationary wave $u_p=100$~m/s obtained in the moving window method using the TKC-MUSCL-SPH method. The profiles were averaged from 75 instantaneous profiles of the 1D SPH approximation (throughout the insertion of 9 pores into the computational domain). The averaged profiles are shown as solid lines, while the dashed lines represent the instantaneous profiles to demonstrate the heterogeneity of the values in the material.}
 	
 \end{figure}

Figure~\ref{fig:profiles} shows the one-dimensional profiles of the longitudinal stress component $-\sigma_{xx}$ and density. The profiles are obtained by averaging 75 instantaneous profiles obtained during the insertion of 9 pores into the computational domain.

\section*{Conclusion}

A modification of the Godunov-type SPH contact method with reconstruction of values at the contact of the MUSCL type and kernel correction (TKC-MUSCL-SPH) is proposed. As test calculations have shown, the proposed method has low numerical diffusion compared to the original CSPH method and provides good accuracy of spatial approximation. Due to the simplicity of implementation and its purely Lagrangian nature, this method is well suited for mesomechanical modeling of shock wave propagation through porous media.

An adaptive moving observation window method for modeling stationary shock waves has been developed. The method is developed to predict shock Hugoniot of porous materials in the absence of experimental data.
We have achieved low amplitude shock wave modeling, in particular, the structure of the stationary wave front of plastic compacting porous copper is investigated.
In this work, we obtained the shock Hugoniot of porous copper at low amplitudes by economically using computational resources, which is a distinctive quality of the proposed method and can be used in particle method simulations to achieve higher resolution than simulations in a shock-related reference frame.  In particular, the simulations allowed us to study the structure of weak shock waves below the kink in the shock Hugoniot of porous copper, which is caused by the complete closure of the pores in the plastic compacting wave.


\end{document}